\pdfoutput=1 
\documentclass{JINST}

\usepackage{graphicx}
\usepackage{epstopdf}
\usepackage{dcolumn}
\usepackage{bm}

\usepackage{amsmath, amssymb, amsthm, amsfonts}
\usepackage{indentfirst}
\usepackage{geometry}

\usepackage[caption=false]{subfig}

\newcommand{\be}{\begin{equation}}
\newcommand{\en}{\end{equation}}

\title{Simulation of plasma loading of high-pressure RF cavities}

\author{K. Yu$^a$,
R. Samulyak$^{a,b}$\thanks{Corresponding author.}~, K. Yonehara$^c$ and B. Freemire$^d$\\
\llap {$^a$}Computational Science Initiative, Brookhaven National Laboratory, \\
Upton, New York 11973, USA\\
\llap {$^b$}Department of Applied Mathematics and Statistics, Stony Brook University, \\
Stony Brook, New York 11794, USA \\
\llap {$^c$}Fermi National Accelerator Laboratory,\\
 Batavia, Illinois 60510, USA\\
\llap {$^d$}Northern Illinois University, \\
DeKalb, IL 60115, USA\\
E-mail: kwangmin.yu@stonybrook.edu}

\abstract{Muon beam-induced plasma loading of radio-frequency (RF) cavities filled with high pressure hydrogen gas with 1\% dry air dopant has been studied via numerical simulations.
The electromagnetic code SPACE, that resolves relevant atomic physics processes, including ionization by the muon beam, electron attachment to dopant molecules, and
electron-ion and ion-ion recombination, has been used. Simulations studies have been performed in the range of parameters typical for practical muon cooling channels.}


\begin{document}

\section{Introduction}

Muon cooling is a challenging problem, critical for the successful design of a muon collider or a neutrino factory.
Ionization cooling is the only viable method to quickly cool the beam temperature within the muon lifetime~\cite{Neuffer04}.
As muons propagate through a material with strong magnetic focusing, their kinetic energy is reduced via ionization processes and
then recovered by RF cavities. Better cooling efficiency is obtained with higher RF gradient. However, the achievable RF gradient is limited by the presence of a static
magnetic field in a vacuum cavity because the density of dark current, which is a seed of electric breakdown in an RF cavity, is increased by magnetic focusing~\cite{Moretti05}.
A high-pressure hydrogen gas-filled RF (HPRF) cavity \cite{Johnson04} has the potential to resolve this problem. Hydrogen gas in the cavity
serves a dual role: it buffers the dark current and serves as the ionization material for the cooling process. A novel ionization cooling channel using the dual function cavity
was proposed and its high cooling efficiency has been demonstrated via simulations~\cite{Derbenev05, Yonehara06}.

A parallel particle-in-cell (PIC) code SPACE, designed for simulations of electromagnetic fields, relativistic particle beams, and plasma chemistry \cite{MOPMN012} has been used
in this work for the simulation of HPRF cavities interacting with muon beams. The atomic physics module of SPACE was validated
in our previous work \cite{YuSam17}.  The validation data were provided by
a series of experiments performed in the Mucool Test Area (MTA) at Fermilab~ \cite{Chung et al,Freemire_thesis,Freemire16} using the HPRF cavity interacting with intense proton beams. In particular, the plasma loading effect has been investigated~\cite{Freemire16}.
Plasma loading takes place when a beam-induced plasma in the HPRF cavity absorbs the stored energy of the electromagnetic field in the cavity and
causes a reduction of the electric field. The plasma loading depends on the amount and mobility of charged particles in the gas.
To decrease the plasma loading effect, adding a small amount of electronegative dopant to the hydrogen gas was proposed and successfully tested experimentally.
Simulations achieved very good agreement with experiments. Comparison of simulations and experiments has also led to the calibration of
some parameters characterizing atomic processes, in particular recombination and attachment rates, which could not be reliably measured in experiments.

In this work, we report simulation results of the interaction of muon beams with two HPRF cavities filled with hydrogen gas at pressures of 100 atm and 160 atm, and containing 1 \% dry air dopant. Simulations support the design of a practical muon ionization cooling channels.

\section{Governing equations and numerical algorithms}
\label{Atomic_Physics}

\subsection{Plasma formation}\label{IONIZATION}

As an intense proton or muon beam propagates in the HPRF cavity, plasma is created by ionization of hydrogen gas molecules due to collisions with beam particles.
Some generated electrons have enough kinetic energy to cause secondary ionization.
The process is described as
\begin{align}
\label{eqno:IONIZATION:1}
\notag p + H_2 &\rightarrow p + {H_2}^+ + e^- ,\\
e^- + H_2 &\rightarrow {H_2}^+ + 2 e^- ,
\end{align}
where $p$, $H_2$, and $e$ are a beam particle (proton or muon), hydrogen molecule, and electron, respectively.

The number of electron-ion pairs $N_{pairs}$, produced in an elementary volume $dV$ of the cavity during time $dt$,
is estimated based on the stopping power of the beam in hydrogen gas
\begin{align}
\label{eqno:IONIZATION:2}
N_{pairs} = \frac{N_b}{W_i}  \frac{dE}{dx} \rho h .
\end{align}
Here ($dE/dx$) is the normalized stopping power of a beam particle per unit density of the absorber material,
$\rho$, $h$, $W_i$, and $N_b$ denote the gas mass density, length of the volume $dV$ along the beam path, average ionization energy, and the number of beam particles
that traverse $d V$ in time $dt$,  respectively.
In Eq. (\ref{eqno:IONIZATION:2}), $W_i$ accounts for both processes described in (\ref{eqno:IONIZATION:1}). The Bethe-Bloch formula \cite{Neuffer04} is used for
calculating the stopping power.

At high pressures over $20$ atm, typical for the HPRF cavity, the $H_2^+$ ions quickly form other clusters of hydrogen:
\begin{align}
\label{eqno:IONIZATION:3}
\notag H_2^+ + H_2 &\rightarrow {H_3}^+ + H, \\
H_{j-2}^+ + 2 H_2 &\rightleftharpoons {H_j}^+ + H_2 ~ (j=5, 7, 9, \ldots).  
\end{align}

\subsection{Recombination}\label{RECOMBINATION}

In the HPRF cavity filled with pure hydrogen gas, electrons created by ionization recombine with hydrogen clusters through either binary or ternary recombination processes:
\begin{align}
\label{eqno:RECOMBINATION:2}
\notag {e}^- + {H_j}^+ &\rightarrow neutrals, \\
{e}^- + {H_j}^+ + H_2 &\rightarrow neutrals + H_2,
\end{align}
where $j = 3, 5, 7, \cdots$.
At the HPRF conditions, ternary recombination is dominant \cite{Chung et al}.
The time evolution of the electron number density is given by the following equation
\begin{equation}
\label{eqno:RECOMBINATION:3}
\frac{d n_e}{d t} = \dot{N_e} - \sum_j \beta_j n_e n_{{H_j}^+},
\end{equation}
where $j = 3, 5, 7, \cdots$, and $n_e$, $n_{{H_j}^+}$, $\dot{N_e}$, and $\beta_j$ denote the number density of electrons, the number density of ${H_j}^j$ hydrogen ion clusters, the production rate of electrons, and the recombination rate of
electrons and ${H_j}^+$, respectively.
Relative populations of hydrogen clusters were not measured in the MTA experiments, and therefore an effective recombination rate $\beta_e$ for an averaged hydrogen ion cluster was used in our models and simulations.
The averaged hydrogen ion cluster that represents the
sum  $\sum_j \beta_j n_{{H_j}^+}$ is denoted as $\beta_e n_{H^+}$.
The effective recombination rate was measured in the MTA experiments at the equilibrium state of plasma (i.e. $d n_e / d t =0$).
In our model, we use the following fit for the effective recombination rate, applicable to transient processes in the plasma,
\begin{equation}
\label{eqno:RECOMBINATION:5}
\beta_e = a X^b,
\end{equation}
where $X = E/P $ is the ratio of electric field magnitude to the gas pressure. Values for the parameters $a$ and $b$ were obtained via comparison of HPRF simulations to various experimentally measured quantities, in particular, the plasma loading \cite{Freemire16}, and extrapolated to the range of parameters used in this work.

\subsection{Attachment and ion - ion recombination}\label{attachment}

When an electronegative gas such as oxygen is added to the hydrogen  gas, a three-body attachment process takes place in the plasma, which is significantly
faster than the electron - ion recombination \cite{Freemire_thesis}.
The negative ions produced by the attachment process recombine with hydrogen ions. The governing equations are
\begin{align}
\label{eqno:ATTACHMENT:1}
\notag \frac{d n_e}{d t} &= \dot{N_e} - \beta_e {n_e} n_{H^+} - \frac{n_e}{\tau},\\
\frac{d n_{H^+}}{d t} &= \dot{N_e} - \beta_e n_e n_{H^+} - \eta n_{H^+} n_{O_2^-},\\
\notag \frac{d n_{O_2^-}}{d t} &= \frac{n_e}{\tau} - \eta n_{H^+} n_{O_2^-},
\end{align}
where $\tau$, $\eta$, and $n_{O_2^-}$ are  the attachment time, effective ion - ion recombination rate, and the number density of dopant gas ions, respectively.

As the recombination and attachment rates depend on  the electric field, they are functions of spatial coordinates and time.
The attachment time and the ion - ion recombination rate have been measured experimentally, but only over a narrow range of RF field amplitudes. Parameters for current simulations were obtained by extrapolation of experimental data and previous simulations. 

\subsection{Numerical algorithms and their implementation}

Simulations of processes in the HPRF cavity have been performed using a parallel particle-in-cell (PIC) code SPACE  that resolves electromagnetic fields, relativistic particle beams, and plasma chemistry \cite{MOPMN012}. For electrostatic problems, SPACE uses a highly adaptive Particle-in-Cloud method \cite{AP-Cloud}.  A set of verification test problems for SPACE is described in \cite{Yu_thesis}. The main feature of SPACE, relevant to
HPRF simulations, is its ability to support  complex atomic physics transformations such as those described in the previous section.
Details of plasma chemistry algorithms as well as a time stepping algorithm that maintains different time scales for the particle beam and plasma can be found in \cite{YuSam17}. 

While the dynamics of the proton beam and plasma were explicitly resolved in simulations, a direct simulation of plasma loading was not feasible due to its non-local nature. In simulations,
the change of the RF field due to plasma loading can be approximated by an RLC resonant circuit formula \cite{Freemire16}. In particular, the RF power dissipation
can be described as
\begin{equation}
\label{eqno:PLASMA LOADING:2}
P = \frac{V(t) [V_{\textrm{max}} - V(t)]}{R} - C V(t) \frac{d V(t)}{d t},
\end{equation}
where $P$, $R$, $C$, $V_{\textrm{max}}$, $V(t)$ are the RF field power, the shunt impedance, the cavity capacitance, the magnitude of the RF voltage, and the
instantaneous voltage value, respectively \cite{Chung et al}.
In the simulations, the total power of plasma loading is computed at each time step as a sum over all computational nodes $N$ of the PIC mesh
\begin{equation}
P(t) = \sum_{j=1}^{N} dw_j(t),
\label{Power}
\end{equation}
where
\begin{equation}
\label{eqno:PLASMA LOADING:1_1}
dw_j(t) = q \int_0^t \left(p_e \mu_e + \mu_+  + (1 - p_e) \mu_-\right)  E_0^2 \textrm{sin}^2 (\omega \tau) d\tau
\end{equation}
is the average energy loss by plasma particle pairs per computational node $j$ during one RF cycle \cite{YuSam17}. Here
$E_0(x,y,z) $, and $\omega$ denote the local amplitude and the
angular frequency of the RF field, respectively, $p_e$ is a fraction of electron in an effective electron-ion pair, and
$\mu_e$, $\mu_+$, and $\mu_-$ denote
mobilities of electrons, and positive and negative ions. An effective electron-ion pair contains one positive ion,
a fraction of an electron $p_e$, and the corresponding fraction of a negative ion $1-p_e$. In pure hydrogen
gas, $p_e = 1$.
The value of $P(t)$ is used to compute the RF field voltage $V(t)$ by integrating the following ordinary differential equation
\begin{equation}
\frac{d V(t)}{d t} = \frac{V_{\textrm{max}} - V(t)}{R C} - \frac{P(t)}{CV(t)}.
\label{ODE}
\end{equation}

Another factor contributing to the plasma loading simulations is the distribution of the RF field ($E_0(x,y,z) $) in the cavity, studied in \cite{Freemire_thesis}.
The RF field amplitude significantly changes in the longitudinal direction, while the change in the radial direction is negligibly small across the plasma column.  Therefore, the RF field amplitude was approximated in the SPACE code as a quadratic polynomial in the longitudinal coordinate, using data from global electromagnetic simulations of the cavity.

\section{Simulation results}\label{SIMULATION}

\subsection{Summary of validation studies}

The SPACE code atomic physics module has been validated in \cite{YuSam17} using results of the Fermilab MTA experiments \cite{Freemire16} on the interaction of proton beams
with high-pressure gas in the HPRF cavity.
Dominant effects of the plasma dynamics in the HPRF cavity have been quantified in numerical simulations and previous analytical  studies.
They showed that ionization electrons are immediately thermalized by interacting with neutral particles and follow the conventional electron transport model.
As reported in the experimental paper~\cite{Chung et al}, electron mobility is significantly reduced by a pressure effect.
Electron capture by an electronegative dopant is also explained by the conventional three body model.
It concludes that the electron capture time can be much shorter than a nanosecond.

Simulations show a very strong reduction of the RF field magnitude in equilibrium for pure hydrogen plasma:
the field is reduced by a factor of 7 at the pressure of 100 atm, and by the factor of 44 at 20.4 atm.
The larger reduction of the electric field at low pressure is due to smaller recombination rates and, therefore,  higher electron density.
If a 1 \% dry air dopant is added to  the hydrogen gas, the reduction of the RF field is greatly mitigated: the reduction factor at 20.4 atm is only 1.7.
Simulations have achieved very good agreement with experiments on plasma loading and related processes.

Simulations also contributed to a better understanding of plasma properties.
In a series of simulations and their comparison with experimentally measured quantities characterizing the plasma loading process, several uncertain properties of the plasma, such as effective recombination rates and the attachment time of electrons to dopant molecules, have been quantified and accurate fit functions for these quantities, valid over a wide range of electric field values and the same pressure values, have been found.

\subsection{Plasma loading of HPRF cavities by muon beams}

In this section, we discuss simulations of the muon beam interaction with HPRF cavities in the parameter space of practical ionization cooling channels using the
validated code SPACE. The main physics parameters  are described in Table \ref{tab:pure_1470}.
\begin{table}
\begin{center}
\begin{tabular}{|l|c|c|}
    \hline
    Parameters & Regime I & Regime II \\ \hline
    Beam kinetic energy & \multicolumn{2}{c|}{120 MeV }\\
    Bunch length & \multicolumn{2}{c|}{3 cm }\\
    Bunch radius & 1 cm & 0.1 cm\\
    Population per  bunch & \multicolumn{2} {c|}{ $5 \times 10^{10}$ }\\
    Bunch spacing & \multicolumn{2}{c|}{3 ns } \\
    Number of bunches & \multicolumn{2}{c|}{20 } \\
    dE / dx & \multicolumn{2}{c|}{4.49 $\textrm{MeV} \textrm{cm}^2 / \textrm{g}$ } \\
    Average ionization energy & \multicolumn{2}{c|}{36.2 eV } \\
    Cavity length & 10 cm & 5 cm \\
   $H_2$ gas pressure & \multicolumn{2}{c|}{ 100 atm and 160 atm  with 1\% dry air}\\
     RF field magnitude &  \multicolumn{2}{c|}{20 MV/m } \\
    Frequency & 325 MHz & 650 MHz\\
    \hline
\end{tabular}
\end{center}
\caption{Parameters of HPRF cavity.}\label{tab:pure_1470}
\end{table}

Simulations model two HPRF cavities: the first one is 10 cm long, with 20 MV/m, 325 MHz RF field, and the second cavity is 5 cm long, with 20 MV/m, 650 MHz RF field. Both cavities are filled with hydrogen gas containing 1\% of dry air dopant. Each cavity was studied at 100 atm and 160 atm of gas pressure. HPRF cavities interact with muon beams consisting of 20 bunches with the length of 3 cm, arriving with the interval of 3 ns.  The muon beam in the first cavity is very wide in the transverse direction, with the radius of 1 cm. In the second cavity, it is focused to the radius of 0.1 cm.
For simulations reported in this paper, we use the following approximations for coefficients determining the rate of atomic processes: the effective electron-ion recombination rate $\beta_e = 1.5\times 10^{-10} X^{-1.2}$, the ion-ion recombination rate $\eta = 1.6\times 10^{-10} X^{-1.0}$, and the electron attachment time to dopant molecule $\tau = 4.0\times 10^{-7} X^{1.0}$, where the ratio of the electric field magnitude to the gas pressure, $X = E/P$, is computed in units of $MV/(m \, psi)$.

\begin{figure}[htp]
\centering
\includegraphics[width=0.5\textwidth]{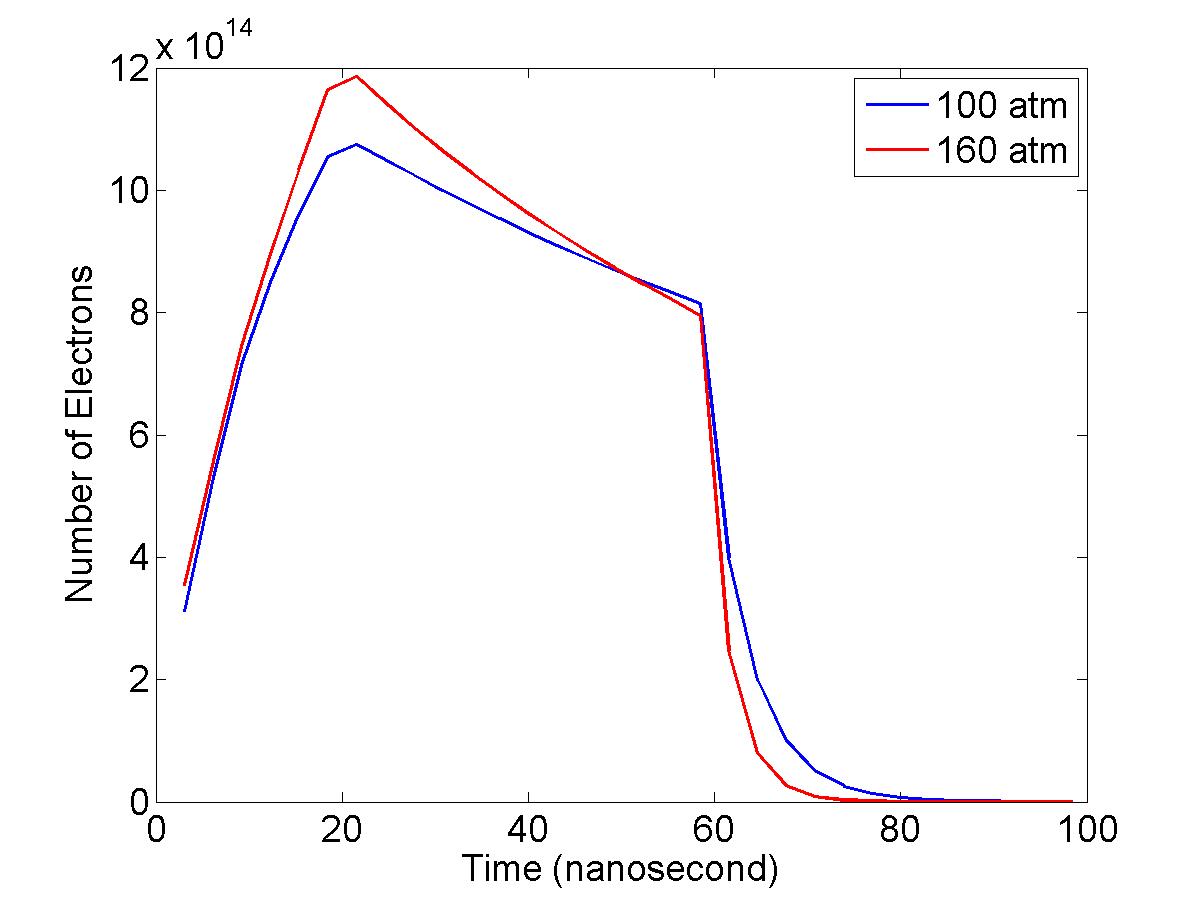}
\caption{Evolution of the number of electrons in the 10 cm long cavity with 325 MHz RF field interacting with the wide muon beam.}
\label{fig:el_10cm_325MHz}
\end{figure}

\begin{figure*}[htp]
\centering
\subfloat [Power dissipated by plasma] {
    \includegraphics[width=0.5\textwidth] {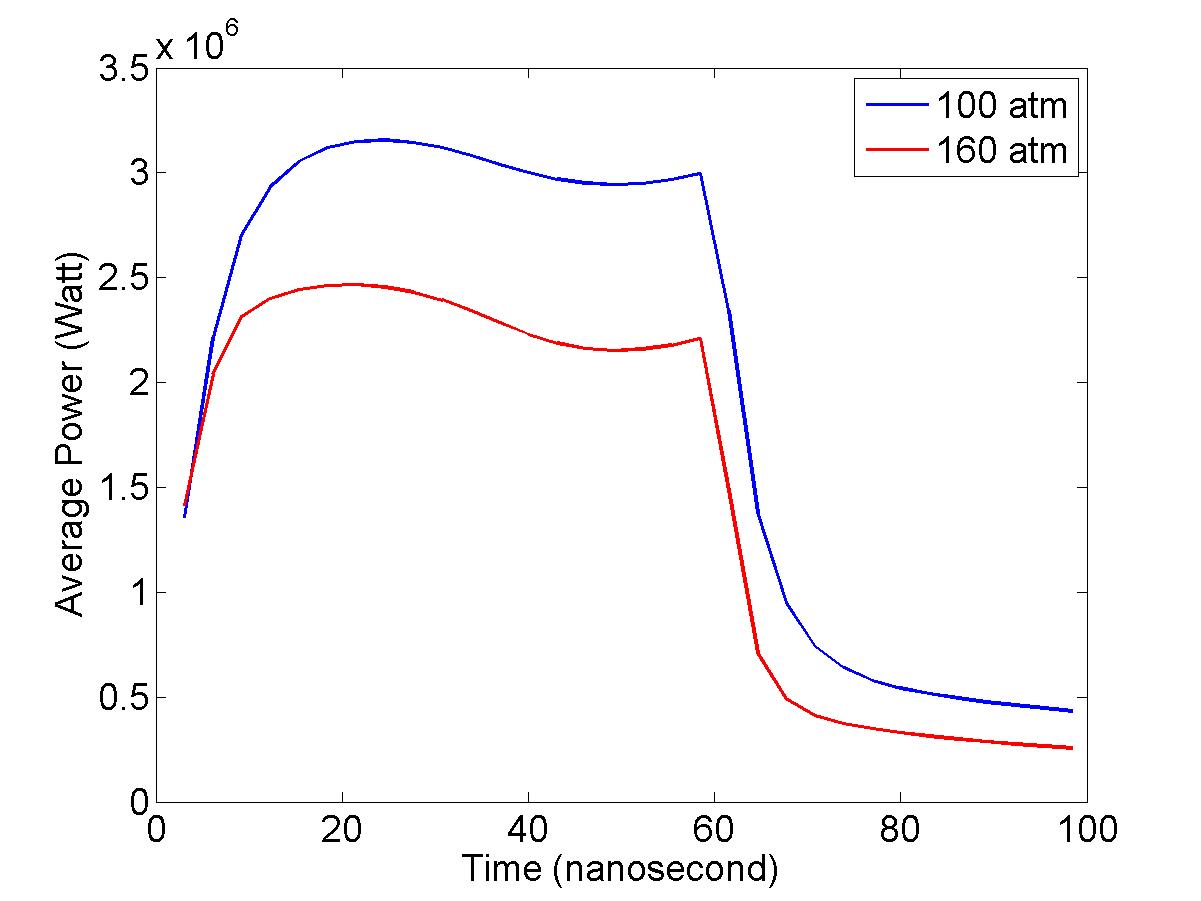}
    \label{fig:power_10cm_325MHz}
}
\subfloat [RF field magnitude] {
    \includegraphics[width=0.5\textwidth] {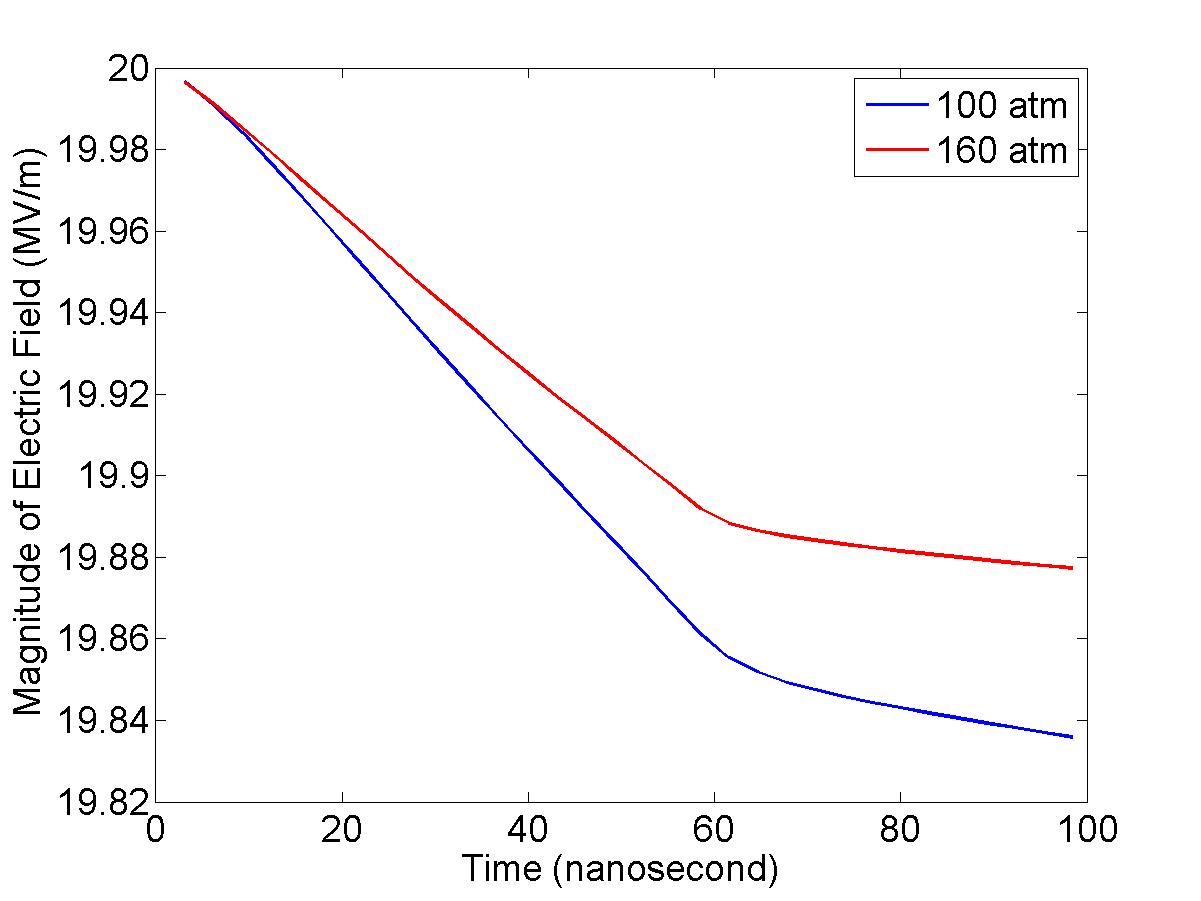}
    \label{fig:E_10cm_325MHz}
}
\caption{Evolution of (a) power dissipated by plasma, averaged  in time over one RF cycle, and (b) RF field magnitude in the 10 cm long cavity with 325 MHz RF field interacting with the wide muon beam. }
\label{fig:10cm_325MHz}
\end{figure*}

Evolution of the total number of electrons in the first cavity is shown in Figure \ref{fig:el_10cm_325MHz}. During first 20 ns, the number of electrons rapidly increases due to ionization, reaching maximum values of  $~1.08\times 10^{15}$ and  $~1.2\times 10^{15}$  at the gas pressures of 100 atm and 160 atm, respectively. With the increase of the electron density, the electron-ion recombination and electron attachment rate to dopant molecules intensifies, overpowers the gas ionization rate by the muon beam, and the number of electrons gradually reduces to  $~0.8\times 10^{15}$ for both hydrogen gas pressures after the arrival of the last muon bunch. After that, the electron density drops to zero within 10 ns.  Evolution of the average power stored in the plasma is shown in Figure \ref{fig:power_10cm_325MHz}.  It reaches maximum values of $~3.3\times 10^6$ and $~2.4\times 10^6$ at the peak of the electron density, slightly reduces towards the end of the beam time, and then rapidly desreases with the decline of the electron density. However, it does not go to zero at 80 ns because of energy stored in positive hydrogen ions and negative dopant ions. As the density of ions slowly deceases by recombination processes, the averaged stored power slowly changes with microsecond  characteristic time scales. The average power stored in plasma at 160 atm is smaller compared to  that at 100 atm despite a slightly larger total number of electrons and ions. This is mostly due to the reduction of electron mobility at higher gas pressure: the electron mobility is approximately 40 \% smaller at 160 atm compared to 100 atm.

The power stored in plasma explains the behaviors of the RF field magnitude, depicted in Figure \ref{fig:E_10cm_325MHz}. The RF field magnitude drops almost linearly during the beam time, and then continues to decrease slightly because of the energy stored in ions. The total decrease of the RF field magnitude is very small - less than 1 \%, which is significantly different compared to large values due to plasma loading of HPRF cavities interacting with proton beams, observed in Fermilab MTF experiments \cite{Freemire16} and simulations \cite{YuSam17}.

The overall time-averaged dynamics of the number of electrons, averaged power stored in plasma, and the RF field magnitude is qualitatively similar for the second HPRF cavity. The  short-time evolution, however, is different, and needs to be explained.

\begin{figure}[htp]
\centering
\subfloat [100 atm] {
\includegraphics[width=0.5\textwidth]{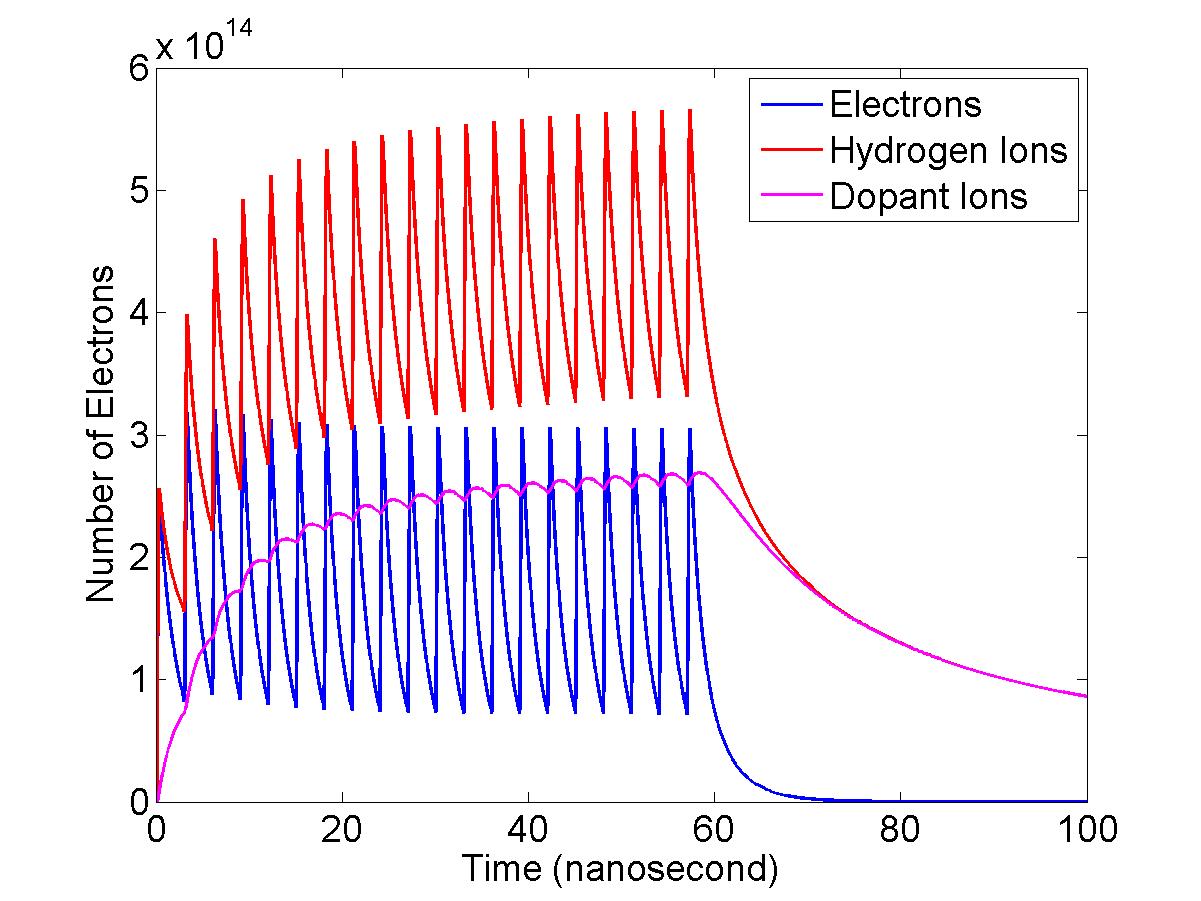}
    \label{fig:electrons_both_p}
    }
\subfloat [160 atm] {
\includegraphics[width=0.5\textwidth]{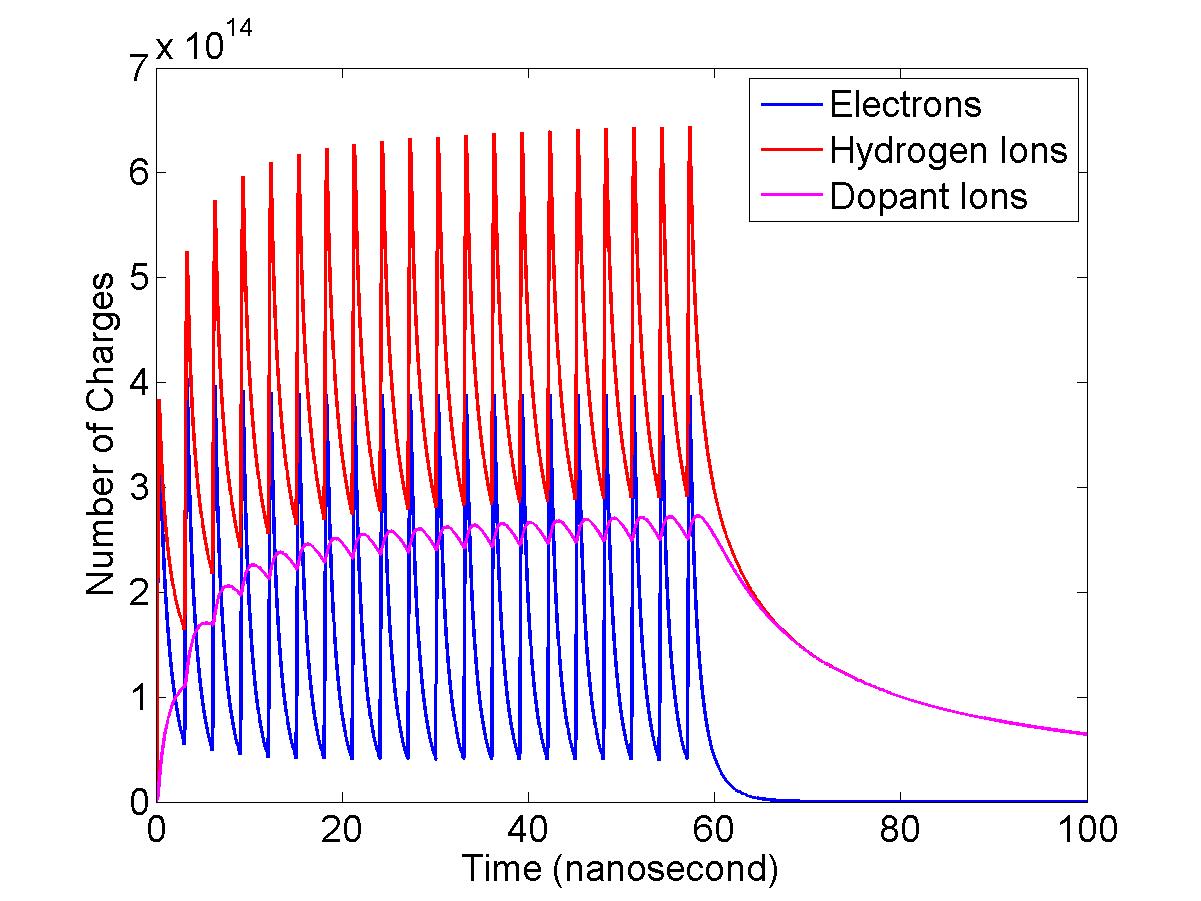}
    \label{fig:charges_100atm}
    }
\caption{Evolution of the number of electrons and positive and negative ions  at (a) 100 atm and (b) 160 atm in the 5 cm long cavity with 650 MHz RF field interacting with strongly focused muon beam.}
\label{fig:charges_5cm_650MHz}
\end{figure}

Evolution of the total number of electrons, and positive and negative ions at 100 atm and 160 atm pressure values in the second cavity is shown in Figure \ref{fig:charges_5cm_650MHz}. As the muon beam is strongly focused to the radius of 0.1 cm, the number density of electrons is almost two orders of magnitude higher along the beam axis compared to the first cavity. Such a high number density of electrons significantly increases the electron recombination and attachment rates. Significant fractions of electrons recombine with hydrogen ions and become attached to dopant molecules within nanosecond time intervals between muon bunches. This explains the highly oscillatory behavior of the number of electrons and, correspondingly, positive and negative ions. Because of fast recombination / attachment processes, the maximum number of electrons is much smaller compared to the previous cavity. 

\begin{figure*}[htp]
\centering
\subfloat [Instantaneous plasma power] {
    \includegraphics[width=0.5\textwidth] {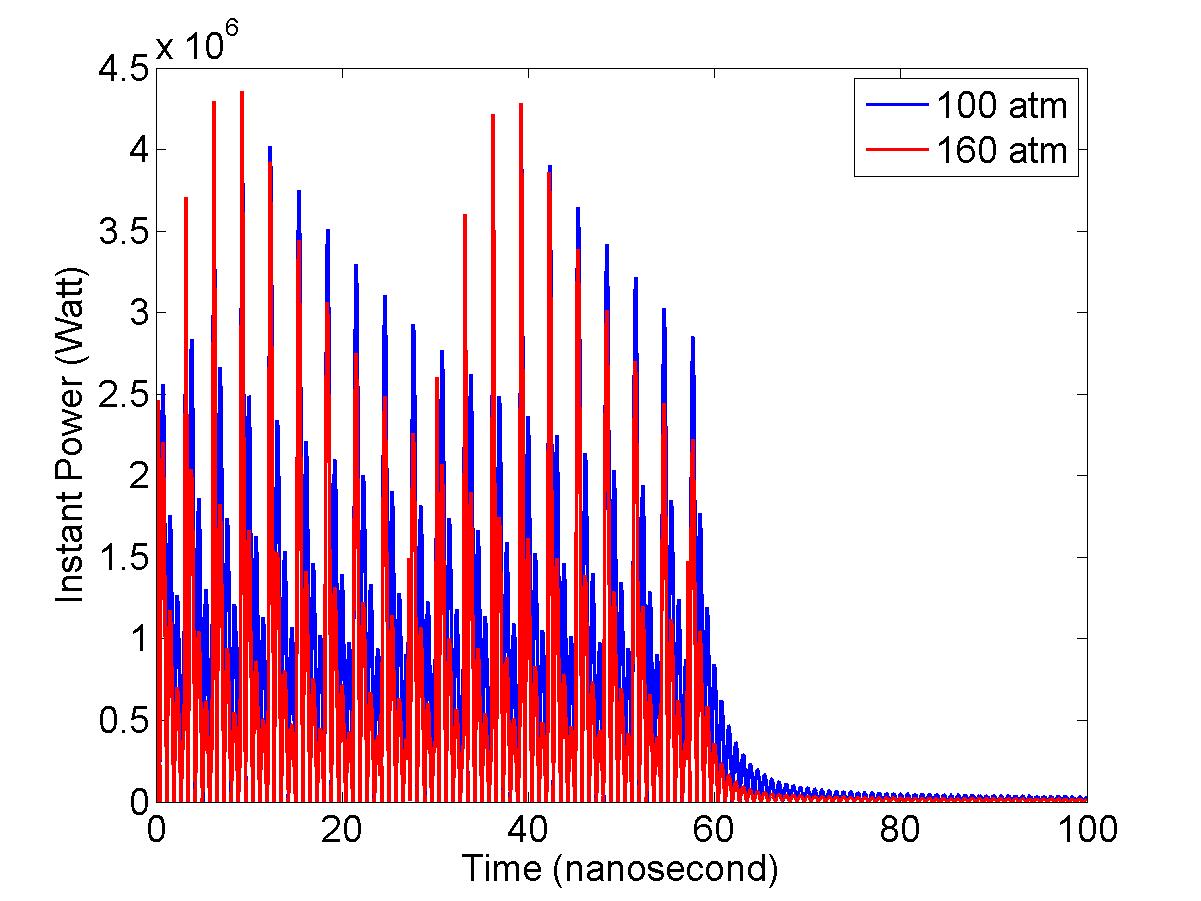}
    \label{fig:instant_power}
}
\subfloat [Power averaged over RF cycle] {
    \includegraphics[width=0.5\textwidth] {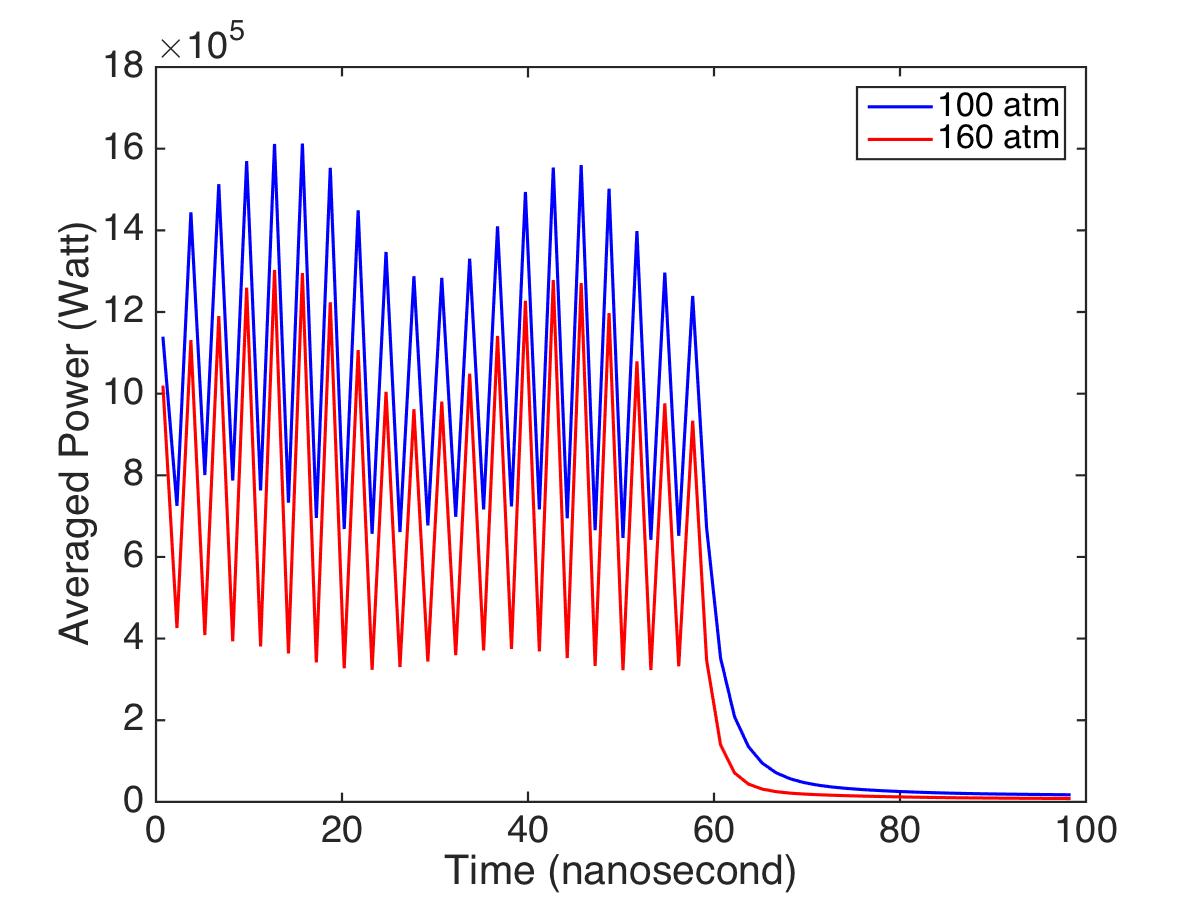}
    \label{fig:averaged_power_rf_cycle}
}
\newline
\subfloat [Power averaged over muon bunch period] {
    \includegraphics[width=0.5\textwidth] {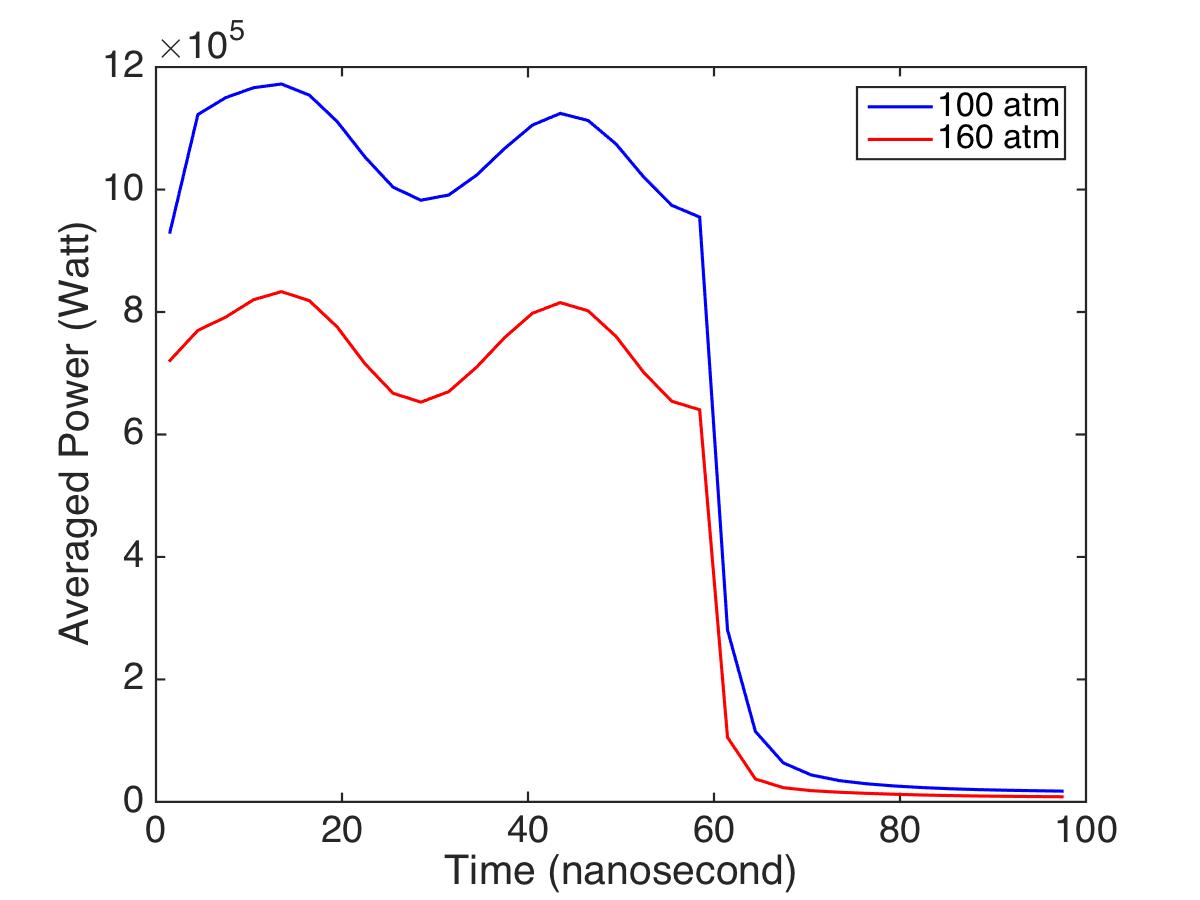}
    \label{fig:averaged_power_beam_cycle}
}
\subfloat [RF field magnitude] {
    \includegraphics[width=0.5\textwidth] {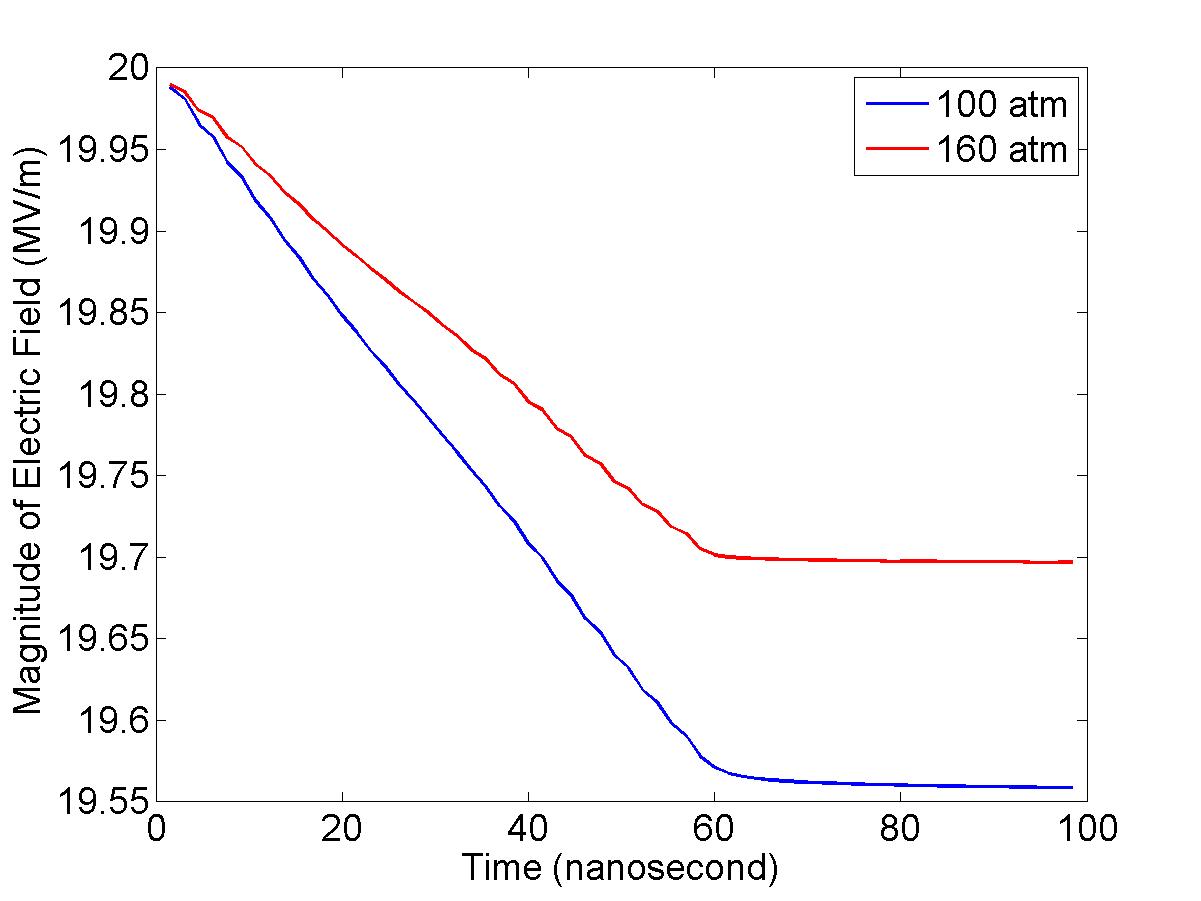}
    \label{fig:E_5cm_650MHz}
}
\caption{Evolution of power dissipated by plasma: (a) instantaneous values of plasma power, (b) plasma power averaged over RF field period,
(c)  plasma power averaged over muon bunch period, and (c) RF field magnitude  in the 5 cm long cavity with 650 MHz RF field interacting with the strongly focused muon beam. }
\label{fig:5cm_650MHz}
\end{figure*}

Similar oscillatory behavior is also evident for the power stored in plasma particles and depicted in Figure \ref{fig:5cm_650MHz}. For
better explanation of the behavior of plasma power, we present the evolution of its instantaneous value in Figure \ref{fig:instant_power}, the evolution of plasma power averaged over the RF field period in Figure  \ref{fig:averaged_power_rf_cycle}, and the evolution of plasma power averaged over the muon bunch period in Figure  \ref{fig:averaged_power_beam_cycle}. The fastest oscillatory behavior of instantaneous plasma power values is caused by fast oscillations of the number density of charges and approximately twice faster 
oscillations of the RF field. Averaged over the RF period, the plasma power oscillates with the frequency of muon bunches. Finally, 
the averaging of plasma power over the muon bunch period results in a smooth variation of plasma power. Lower values of the plasma power
at 28 - 30 $ns$ (see Figure  \ref{fig:averaged_power_beam_cycle}) is the result of small details in the evolution of  plasma particles. In particular, the highest values of the electron number density for both values of gas pressures (see Figure \ref{fig:charges_5cm_650MHz}) are obtained after the arrival of only the second and third muon bunches. The number of dopant ions at this time is still low. But as the electron mobility is significantly larger compared to the 
mobilities of positive and negative ions, the plasma power is close to its maximum value (see Figure \ref{fig:averaged_power_beam_cycle}). After that, the averaged number of negative and positive ions continues to slightly increase while the averaged number of electrons slightly reduces as attachment processes intensify. The more rapid increase of ions is responsible for the increase of the plasma power until approximately 7 ns. After that, further reduction of electrons causes the reduction of the plasma power despite further increase of the number of positive and negative ions: plasma power is much more sensitive to the number electrons as
their mobility is approximately 43 times larger than the mobility of positive hydrogen ions and approximately 71 times larger compared to the mobility of negative ion dopants. The local minimum of plasma power is achieved at ~ 28 - 30 ns. At this time, the number of electrons, averaged over the muon beam period, reaches a steady state due to the balance of ionization, recombination and electron attachment processes. With a constant averaged number of electrons, further increase of the number of ions causes a slight increase in  the plasma power, and a second local maximum is reached at about 42 ns. It then reduces again because of a slight reduction of the electron number. The reduction of the RF field magnitude due to plasma loading, that causes a decrease of mobilities of plasma particles, and a 3D spatial dependence of all quantities add extra complications to the analysis.

The reduction of the RF field amplitude, shown in Figure \ref{fig:E_5cm_650MHz}, is not sensitive to such small scale oscillations. While the reduction of the RF field magnitude is very small in both cases, it is larger in the second cavity despite smaller power stored in plasma compared to the first cavity: the RF power in the second cavity is reduced by 2.2 \% in 100 atm hydrogen gas at 80 ns compared to 0.8 \% in the first cavity. This difference is due to the geometry and parameters of each cavity:  the capacitance and resistance of the 10 cm long, 325 MHz cavity are $6.1\times 10^{-12}$ F  and  8.7 MOhm, respectively, while these parameters for the 5 cm long, 650 MHz cavity are  $3.0\times 10^{-12}$ F  and    3.1 MOhm.

\section{Conclusions}
\label{Conclusion}

Numerical simulations of the interaction of intense muon beams with two  HPRF cavities filled with hydrogen gas at the pressures of 100 and 160 atm and containing 1\% of dry air dopant have been performed using code SPACE. The atomic physics module of SPACE is designed to resolve all relevant processes occurring in the HPRF cavity. They include ionization by muon impact, electron attachment to dopant molecules, formation of cascades of heavier hydrogen ions, electron-ion and ion-ion recombination processes, as well as experimental data-based models of electron and ion mobilities in the dense gas. The atomic physics module has been validated in our previous work \cite{YuSam17} by comparing simulations with experiments in the Fermilab MTA facility which operated HPRF cavities filled with pure hydrogen gas and hydrogen gas containing dry air dopant interacting with intense proton beams. Simulations showed  a very strong reduction of the RF field magnitude in equilibrium for pure hydrogen plasma, much smaller reduction in the presence of dry air dopant, and  achieved very good agreement with experiments.

The validated SPACE code has been used in this work for the study of atomic processes in HPRF cavities and their plasma loading by intense muon beams. The parameter set of numerical simulations is relevant  to the design of practical muon cooling channels. We have analyzed the evolution of atomic processes in HPRF cavities, power stored in plasma, and their influence on the reduction of the RF field magnitude. We show that the plasma dynamics strongly depends on the muon beam focusing: tightly focused beams create high electron number density which in turn intensifies the electron attachment to dopant molecules and causes a rapid reduction of the number of free electrons and its oscillation with the frequency of muon  bunches. The reduction of the RF field amplitude, however, is not sensitive to these fast oscillations. The reduction of the RF field amplitude was negligibly small for all  cases studied in this work. Several other issues, not unique to muon calling channels, affect the HPRF cavity performance. One of them is the beam loading which significantly affects the HPRF cavity accelerating mode for high intensity muon beams \cite{ChungIPAC13}. 

\vskip3mm
{\bf Acknowledgement.}
This research has been partially supported by the DOE Muon Accelerator Program.
This manuscript has been authored in part by Brookhaven Science Associates, LLC, under Contract No.
DE-SC0012704 with the US Department of Energy.

\appendix
\section{Derivation of RLC Circuit Formula for HPRF Cavity Plasma Loading}

\begin{figure}[htp]
\centering
\includegraphics[width=0.9\textwidth]{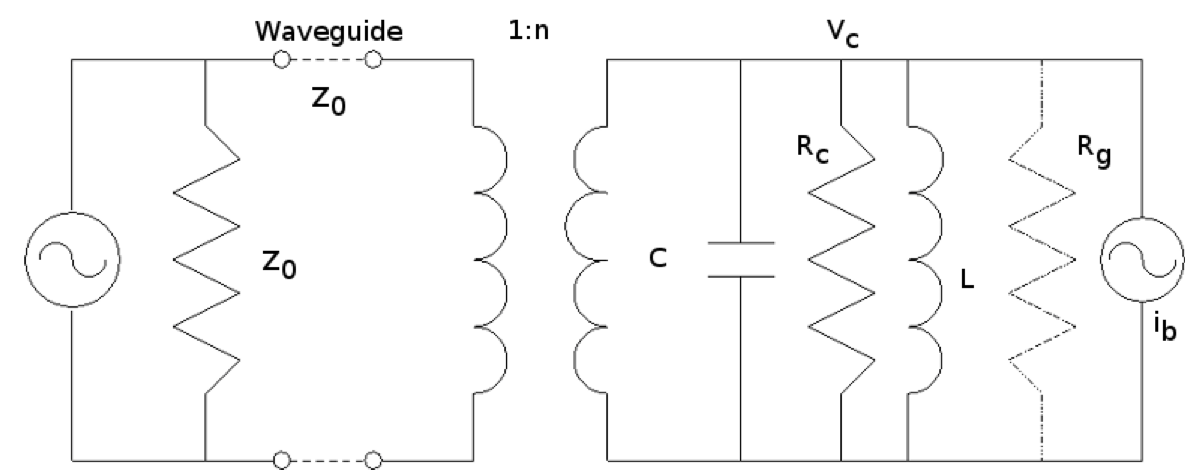}
\caption{Circuit diagram for an RF cavity experiencing beam and plasma loading.  The generator provides a voltage, V$_{0}$ with a characteristic impedance, Z$_{0}$, and is coupled to the cavity through a waveguide, which acts as a transformer.  The cavity voltage, resistance, capacitance, and inductance are given by V$_{c}$, R$_{c}$, C, and L, respectively.  The plasma provides an additional resistance, R$_{g}$, and the beam acts as a current source, i$_{b}$.}
\label{fig:CircuitDiagram}
\end{figure}

The plasma generated by a beam ionizing the gas within the cavity is an additional current path, and therefore acts as an additional resistance.  The cavity will transfer stored energy to the plasma, where the power delivered is the time rate of change of the stored energy in the cavity:

\begin{equation}
P_{c}[t] = \frac{d}{dt} \left( \frac{1}{2} \mathrm{C} V[t]^{2} \right) = \mathrm{C} V[t] \frac{dV[t]}{dt}.
\label{eq:PowerCavity}
\end{equation}

The cavity continues to receive power from the generator, in this case a klystron, given by:

\begin{equation}
P_{in} = \frac{1}{2} V I^{*} = \frac{1}{2} Z_{in} |I|^{2} = \frac{1}{2} |V|^{2} \frac{1}{Z_{in}^{*}}
		= \frac{1}{2} |V|^{2} \left( \frac{1}{\mathrm{R}} + \frac{j}{\omega \mathrm{L}} - j \omega \mathrm{C} \right)
\label{eq:PowerIn}
\end{equation}

At resonance, $Z_{in} = R$, and Equation \ref{eq:PowerIn} reduces to:

\begin{equation}
P_{in} = \frac{1}{2} \frac{|V|^{2}}{\mathrm{R}} = \frac{1}{2} I V
\label{eq:PowerInSimplified}
\end{equation}

For a matched load (cavity) and generator, Z$_{0}$ = R$_{c}$ in Figure \ref{fig:CircuitDiagram}.  The power coming from the source (klystron) is as a function of time is:

\begin{equation}
P_{k}[t] = \frac{1}{2} I[t] V[t] = \frac{1}{2} \left( \frac{\mathrm{V}_{0} - V[t]}{\frac{1}{2}\mathrm{R}_{c}} \right) V[t]
	= \frac{(\mathrm{V}_{0} - V[t]) V[t]}{\mathrm{R}_{c}}
\label{eq:PowerKlystron}
\end{equation}

The power being delivered to the gas is then the difference between what the source is supplying the cavity, and what the cavity is feeding the gas:

\begin{equation}
P_{g}[t] = P_{k}[t] - P_{c}[t] = \frac{(\mathrm{V}_{0} - V[t]) V[t]}{\mathrm{R}_{c}} - \mathrm{C} V[t] \frac{dV[t]}{dt}
\label{eq:PowerGas}
\end{equation}

Note that at early times, V$_{0} \approx V$, and the majority of power delivered to the gas comes from the stored energy of the cavity, i.e. the second term in Equation \ref{eq:PowerGas} is larger than the first term.  As the cavity discharges $V$ and $dV/dt$ become small, and the majority of power delivered to the gas comes directly from the klystron, i.e. the first term is larger than the second.


\end{document}